\magnification=1200
\def\qed{\unskip\kern 6pt\penalty 500\raise -2pt\hbox
{\vrule\vbox to 10pt{\hrule width 4pt\vfill\hrule}\vrule}}
\centerline{PERTURBATION THEORY FOR LYAPUNOV EXPONENTS OF A TORAL MAP:}
\centerline{EXTENSION OF A RESULT OF SHUB AND WILKINSON.}
\bigskip
\centerline{by David Ruelle\footnote{*}{Mathematics Dept., Rutgers University, and IHES.  91440 Bures sur Yvette, France.\break $<$ruelle@ihes.fr$>$}.}
\bigskip\bigskip\noindent
	{\leftskip=2cm\rightskip=2cm\sl Abstract.  Starting from a hyperbolic toral automorphism, we obtain, for a small volume preserving perturbation, an exact and rigorous second order perturbation expansion of the Lyapunov exponents.\par}
\vfill\eject

	We consider volume preserving perturbations $F$ of a diffeomorphism $F_0=(\Phi,J)$ of ${\bf T}^{m+1}={\bf T}^m\times{\bf T}$, where $\Phi$ is a hyperbolic automorphism of ${\bf T}^m$, and $J$ is a translation of ${\bf T}$.  Writing $F=F_0+aF'$, we shall show that the Lyapunov exponents for $(F,{\rm volume})$ can be expanded to second order in $a$ (Theorem 1).  In particular, the central Lyapunov exponent $\lambda^c$ of $(F,{\rm volume})$, to second order in $a$, is generally $\ne0$ (Corolary 11).  For a special family of perturbations one obtains particularly simple formulae, first noted by Shub and Wilkinson [10].  We recover their result in Theorem 12.  We deviate from [10] mostly in that we don't have differentiability of $\lambda^c$, only a second order expansion around $a=0$.  The ideas used here are largely those in Shub and Wilkinson [10], and can be appreciated in the background provided by Hirsh, Pugh and Shub [6], Burns and Wilkinson [3], Ruelle and Wilkinson [9].  For recent work related to Lyapunov exponents, see also Bonatti, G\'omez-Mont and Viana [2], Avila and Bochi [1].
\medskip
	After completing the writing of this paper, the author received a preprint by D. Dolgopyat [4], which develops similar ideas in a more general setting, but without the specific formulas we obtain here.
\medskip
	{\bf 1. Theorem.}  
\medskip
	{\sl Let $\Phi$ be a hyperbolic automorphism of ${\bf T}^m$, and $J:y\mapsto y+\alpha$ {\rm (mod 1)} a translation of ${\bf T}$.  Define $F_0=(\Phi,J)$, and let $F=F_0+aF'$ be a $C^2$ perturbation of $F_0$, volume preserving to first order in $a$.  (We take $F':{\bf T}^{m+1}\mapsto{\bf R}^{m+1}$ and $F_0\xi+aF'(\xi)$ has to be understood {\rm (mod 1)} in each component).  Let $\lambda_1<\lambda_2<\ldots$ be the Lyapunov exponents of $(F_0,{\rm volume})$ and $m_1,m_2,\ldots$ their multiplicities (the exponent $=0$ occurs with multiplicity $1$).  Also let $\lambda_a^{(1)}\le\lambda_a^{(2)}\le\ldots$ be the Lyapunov exponents of $(F,{\rm volume})$ repeated according to multiplicity.  Then we have the second order expansion
$$	\sum_{\ell=m_1+\ldots+m_{r-1}+1}^{m_1+\ldots+m_r}\lambda_a^{(\ell)}
	=m_r\lambda_r+a^2L_r+o(a^2)      $$
If $m_r=1$, and writing $\lambda_r=\lambda_0^{(\ell)}$, we have
$$	\lambda_a^{(\ell)}=\lambda_0^{(\ell)}+a^2L^{(\ell)}+o(a^2)      $$
(this applies in particular to $\lambda^c=\lambda_a^{(\ell)}$ for $\lambda_0^{(\ell)}=0$).}
\medskip
	An explicit expression for $L_r$ can be obtained (see Proposition 9).  We do not assume ergodicity of $(F,{\rm volume})$, and therefore we use {\it integrated} Lyapunov exponents (averaged over the volume), see however Remark 15(a).
\medskip
	If one likes one may to $F_0+aF'$ add terms of higher order in $a$ so that the sum $F$ is exactly volume preserving.  These higher order terms will not change our results, and are omitted in what follows.
\medskip
	{\bf 2. Normal hyperbolicity.}
\medskip
	As in [10], we invoque the theory of normal hyperbolicity of [6].  We start from the fact that $F_0$ is normally hyperbolic to the smooth fibration of ${\bf T}^{m+1}$ by circles $\{x\}\times{\bf T}$.  Taking some $k\ge2$ we apply [6] Theorems (7.1), (7.2).  Thus we obtain a C$^1$ neighborhood $U$ of $F_0$ in the C$^k$ diffeomorphisms of ${\bf T}^{m+1}$ such that, for $F\in U$, there is an equivariant fibration $\pi:{\bf T}^{m+1}\to{\bf T}^m$ with 
$$	\pi F=\Phi\pi      $$
The fibers $\pi^{-1}\{x\}$ are C$^k$ circles forming a continuous fibration of ${\bf T}^{m+1}$ (this fibration is in general not smooth).  Furthermore there is a $TF$-invariant continuous splitting of $T{\bf T}^{m+1}$ into three subbundles:
$$	T{\bf T}^{m+1}=E^s+E^u+E^c      $$
such that $E^c$ is 1-dimensional tangent to the circles $\pi^{-1}\{x\}$, $E^s$ is $m^s$-dimensional contracting and $E^u$ is $m^u$-dimensional expanding for $TF$.  
\medskip
	If $\lambda_r<0$ (and $F$ is in a suitable C$^1$-small neighborhood $U$ of $F_0$), we can introduce a continuous vector subbundle $E^r$ of $T{\bf T}^{m+1}$ which consists of vectors contracting under $TF^n$ faster than $(\lambda_r+\epsilon)^n$ where $\epsilon>0$ and $\lambda_r+\epsilon<\lambda_{r+1}$.  In fact $E^r$ is a hyperbolic (attracting) fixed point for the action induced by $TF^{-1}$ on the bundle of $m_1+\ldots+m_r$ dimensional linear subspaces of $T{\bf T}^{m+1}$ (over $F^{-1}$ acting on $T^{m+1}$).
\medskip
	If $\lambda_r>0$, replacement of $F$ by $F^{-1}$ similarly yields a continuous subbundle $\bar E^r$ of $m_r+\ldots$ dimensional subspaces.
\medskip
	{\bf 3. Proposition.}
\medskip
	{\sl Assume that $F$ is of class {\rm C}$^k$, $k\ge2$, and that $F$ is $C^k$ close to $F_0$.  The bundles $E^r$, $\bar E^r$, when restricted to a circle $\pi^{-1}\{x\}$ are of class {\rm C}$^{k-1}$, continuously in $x$.}
\medskip
	If ${\cal G}$ denotes the (Grassmannian) manifold of $m_1+\ldots+m_r$ dimensional linear subspaces of ${\bf R}^{m+1}$, we may identify the bundle of $m_1+\ldots+m_r$ dinensional linear subspaces of $T{\bf T}^{m+1}$ with ${\bf T}^{m+1}\times{\cal G}$.  We denote by ${\cal E}\in{\cal G}$ the spectral subspace of the matrix defining $\Phi$ corresponding to the smallest $m_1+\ldots+m_r$ eigenvalues (in absolute value, and repeated according to multiplicity).
\medskip
	If ${\cal F}_0$ is the action defined by $TF_0$ on $T{\bf T}^{m+1}\times{\cal G}$, the circles $\{x\}\times{\bf T}\times\{{\cal E}\}$ form an ${\cal F}_0$ invariant fibration of ${\bf T}^{m+1}\times\{{\cal E}\}$, to which ${\cal F}_0$ is normally hyperbolic.  If $F$ is C$^k$ close to $F_0$, the corresponding C$^{k-1}$ action ${\cal F}$ is normally hyberbolic to a pertubed fibration where $\{x\}\times{\bf T}\times\{{\cal E}\}$ is replaced by $E^r|\pi^{-1}\{x\}$.  According to [6] Theorem 7.4, Corollary (8.3) and the following Remark 2, we find that the C$^{k-1}$ circle $E^r|\pi^{-1}\{x\}\subset{\bf T}^{m+1}\times{\cal G}$ depends continuously on $x\in{\bf T}^{m+1}$.  Similarly for $\bar E$.\qed
\medskip
	Note that in [10], the C$^r$ section theorem is used in a similar situation, giving estimates uniform in $x$.  However, continuity in $x$ (not just uniformity) will be essential for us in what follows.
\medskip
	{\bf 4. Corollary.}
\medskip
	{\sl The splitting $T{\bf T}^{m+1}=E^s+E^u+E^c$ when restricted to a circle $\pi^{-1}\{x\}$ is of class {\rm C}$^{k-1}$, continuously in $x$.}
\medskip
	It is clear that $E^c|\pi^{-1}\{x\}$ is of class C$^{k-1}$ because it is the tangent bundle to the C$^k$ circle $\pi^{-1}\{x\}$.  As to $E^s$, $E^u$, they are special cases of $E^r$, $\bar E^r$.\qed
\medskip
	{\bf Notation.}
\medskip
	Remember that $F=F_0+aF'$, and fix $F'$.  We shall use the notation $\pi_a$, $E_a^r$, \dots to indicate the $a$-dependence of $\pi$, $E^r$, \dots
\medskip
	{\bf 5. Proposition.}
\medskip
	{\sl For small $\epsilon>0$ there is a continuous function $x\mapsto\gamma_x$ from ${\bf T}^m$ to C$^k({\bf T}\times(-\epsilon,\epsilon)\to{\bf T}^m)$ such that $\gamma_x(y,0)=0$ and $\pi_a^{-1}\{x\}=\{(x+\gamma_x(y,a),y):y\in{\bf T}\}$.}
\medskip
	To see this define $\tilde F:{\bf T}^{m+1}\times(-\epsilon,\epsilon)\to{\bf T}^{m+1}\times(-\epsilon,\epsilon)$ by $\tilde F(\xi,a)=((F_0+aF')(\xi),a)$ and observe that $\tilde F$ is normally hyperbolic to the 2-dimensional manifolds 
$$	\cup_{a\in(-\epsilon,\epsilon)}(\pi_a^{-1}\{x\},a)      $$
and these are thus C$^k$ 2-dimensional submanifolds of ${\bf T}^{m+1}\times(-\epsilon,\epsilon)$.\qed
\medskip
	We may in the same manner replace $\pi_a^{-1}\{x\}$ by $\cup_{a\in(-\epsilon,\epsilon)}(\pi_a^{-1}\{x\},a)$ in Proposition 3 and Corollary 4.  Writing $E_a$ for $E_a^r$, $\bar E_a^r$, $E_a^s$, $E_a^u$, $E_a^c$, we obtain that $(\cdot,a)\mapsto E_a(\cdot)$, when restricted from ${\bf T}^{m+1}\times(-\epsilon,\epsilon)$ to $\cup_{a\in(-\epsilon,\epsilon)}(\pi_a^{-1}\{x\},a)$ is of class C$^{k-1}$.  We rephrase this as follows:
\medskip
	{\bf 6. Proposition.}
\medskip
	{\sl The map 
$$	x\mapsto\{(y,a)\mapsto E_a(x+\gamma_x(y,a),y)\}      $$
where $E_a$ stands for $E_a^r$, $\bar E_a^r$, $E_a^s$, $E_a^u$, $E_a^c$, is continuous ${\bf T}^m\to{\rm C}^{k-1}({\bf T}\times(-\epsilon,\epsilon)\to{\rm Grassmannian}\enspace{\rm of}\enspace{\bf R}^{m+1})$ where we have used the identification $T{\bf T}^{m+1}={\bf T}^{m+1}\times{\bf R}^{m+1}$.}\qed
\medskip
	{\bf Notation.}
\medskip
	From now on we write  $E_a$ for $E_a^r$, $\bar E_a^r$, $E_a^s$, $E_a^u$, $E_a^c$.  When $a=0$, $E_0$ is a trivial subbundle of $T{\bf T}^{m+1}={\bf T}^{m+1}\times{\bf R}^{m+1}$, and we shall write $E_0={\bf T}^{m+1}\times{\cal E}$, denoting thus by ${\cal E}$ a spectral subspace of the matrix on ${\bf R}^{m+1}$ defining $(\Phi,1)$.  We denote by ${\cal E}^\perp$ the complementary spectral subspace.
\medskip
	Taking $k=2$ we have then:
\medskip
	{\bf 7. Corollary.}
\medskip
	{\sl There are linear maps $G(x,y),R(x,y,a):{\cal E}\to{\cal E}^\perp$ such that $G(x,y)$ depends continuously on $(x,y)\in{\bf T}^m\times{\bf T}$, $R(x,y,a)$ on $(x,y,a)\in{\bf T}^m\times{\bf T}\times(-\epsilon,\epsilon)$,
$$	E_a(x+\gamma_x(y,a),y)
	=\{X+aG(x,y)X+R(x,y,a)X:X\in{\cal E}\}      $$
and $||R(x,y,a)||$ is $o(a)$ uniformly in $x,y$.}\qed
\medskip
	Notice now that, if $\tilde x=\pi_a(x,y)$, then $x=\tilde x+\gamma_{\tilde x}(y,a)$, where $\gamma_{\tilde x}(y,a)=O(a)$. Now
$$	E_a(x,y)=E_a(\tilde x+\gamma_{\tilde x}(y,a),y)
	=\{X+aG(\tilde x,y)X+R(\tilde x,y,a)X:X\in{\cal E}\}      $$
differs from 
$$	E_a(x+\gamma_x(y,a),y)
	=\{X+aG(x,y)X+R(x,y,a)X:X\in{\cal E}\}      $$
by the replacement $\tilde x\to x$ in the right-hand side, and since dist$(\tilde x,x)=O(a)$, we find that dist$(E_a(x,y),E_a(x+\gamma_x(y,a),y))=o(a)$.  Therefore, changing the definition of $R$, we can again write:
\medskip
	{\bf 8. Corollary.}
\medskip
	{\sl There are linear maps $G(x,y),R(x,y,a):{\cal E}\to{\cal E}^\perp$, depending continuously on their arguments, such that 
$$	E_a(x,y)=\{X+aG(x,y)X+R(x,y,a)X:X\in{\cal E}\}      $$
and $||R(x,y,a)||$ is $o(a)$ uniformly in $x,y$.}\qed
\medskip
	We may write $T_\xi F=T_\xi(F_0+aF')=D_0+aD'(\xi)$ where $D_0$ does not depend on $\xi$ and preserves the decomposition $T_\xi M={\cal E}+{\cal E}^\perp$.  If we apply $TF$ to an element $X+aGX+RX$ of $E_a$ (as in Corollary 8) we obtain $X_1+$ element of ${\cal E}^\perp\in E_a$, with $X_1\in{\cal E}$:
$$	X_1=D_0X+aD'X+a^2D'GX+aD'RX
	\qquad{\rm projected\enspace on\enspace}{\cal E}\eqno{(1)}      $$
Under $(TF)^\wedge$, the volume element $\theta$ in $E_a(\xi)$ is multiplied by a factor $M(\xi,a)$, and the projection in ${\cal E}$ of $(TF)^\wedge\theta$ is equal to the projection in ${\cal E}$ of $\theta$ multiplied by a factor $N(\xi,a)$ such that
$$	M(\xi,a)=N(\xi,a)+\ell_a(\xi)-\ell_a(F\xi)      $$
for suitable $\ell_a$.  We may compute $N$ from (1):
$$	N(\xi,a)=N_{(0)}+aN_{(1)}(\xi)+a^2N_{(2)}(\xi)+o(a^2)      $$
\bigskip
	To proceed we take now $E_a=E_a^r$, and assume $\lambda_r<0$.  We have then, writing $d\xi$ for the volume element in ${\rm T}^{m+1}$, 
$$	L_a=\sum_{\ell=1}^{m_1+\ldots+m_r}\lambda_a^{(\ell)}
	=\int d\xi\,\log M(\xi,a)=\int d\xi\,\log N(\xi,a)      $$
$$	=L_{(0)}+aL_{(1)}(\xi)+a^2L_{(2)}(\xi)+o(a^2)\eqno{(2)}      $$
More precisely, we shall prove
\vfill\eject
	{\bf 9. Proposition.}
\medskip
	{\sl If $\lambda_r<0$, we have 
$$	\sum_{\ell=1}^{m_1+\ldots+m_r}\lambda_a^{(\ell)}
	=\sum_{k=1}^rm_k\lambda_k+a^2L+o(a^2)      $$
where 
$$	L={1\over2}\sum_{n=-\infty}^\infty\int d\xi\,
	{\rm Tr}_{\cal E}(D_0^{-1}D'(\xi))
	{\rm Tr}_{\cal E}(D_0^{-1}D'(F_0^n\xi))\ge0      $$
and ${\cal E}$ is the spectral subspace of the matrix defining $\Phi$ corresponding to the smallest $m_1+\ldots+m_r$ eigenvalues (in absolute value, and repeated according to multiplicity).}
\medskip
	The proof that $L\ge0$ is postponed to Remark 15(b).
\medskip
	The proposition is obtained by comparing formula (2) with the formula (5) below, which we shall obtain by a second order perturbation calculation.
\medskip
	To first order in $a$ we have 
$$	F^n=(F_0+aF')^n
	=F_0^n+a\sum_{j=1}^nF_0^{n-j}\circ F'\circ F_0^{j-1}      $$
hence
$$	T_\xi F^n=D_0^n+a\sum_{j=1}^nD_0^{n-j}D'(F^{j-1}\xi)D_0^{j-1}      $$
If we apply $TF^n$ to $X+aGX+RX\in E_a$ we obtain $X_n+$ element of ${\cal E}^\perp\in E_a$, with $X_n\in{\cal E}$.  To zero-th order in $a$, $X_n=D_0^nX$, so we may write to first order $X_n=D_0^nX+aY_n(\xi)$.  Therefore, to first order in $a$,
$$	D_0^nX+aY_n(\xi)+aG(F^n\xi)D_0^nX
=D_0^nX+a\sum_{j=1}^nD_0^{n-j}D'(F^{j-1}\xi)D_0^{j-1}X+aD_0^nG(\xi)X      $$
and, taking the components along ${\cal E}^\perp$,
$$	G(F^n\xi)D_0^nX
=\sum_{j=1}^nD_0^{n-j}D'_\perp(F^{j-1}\xi)D_0^{j-1}X+D_0^nG(\xi)X      $$
where $D'_\perp(.)$ is $D'(.)$ followed by taking the component along ${\cal E}^\perp$, or 
$$	\sum_{j=1}^nD_0^{-j}D'_\perp(F^{j-1}\xi)D_0^{j-1}X+G(\xi)X
	=D_0^{-n}G(F^n\xi)D_0^nX      $$
When $n\to\infty$, the right-hand side tends to zero (exponentially fast, remember that $X\in{\cal E}$, $GX\in{\cal E}^\perp$).  Therefore 
$$	G(\xi)X
	=-\sum_{j=1}^\infty D_0^{-j}D'_\perp(F^{j-1}\xi)D_0^{j-1}X      $$
which we shall use in the form
$$	G(\xi)X
=-\sum_{n=0}^\infty D_0^{-n-1}D'_\perp(F_0^n\xi)D_0^nX\eqno{(3)}      $$
where we have written $F_0^n$ instead of $F^n$ since $G$ is evaluated to order 0 in $a$.  (The right-hand side is an exponentially convergent series).
\medskip
	Returning to (1) we see that, to second order in $a$,
$$	X_1=D_0X+aD'(\xi)X+a^2D'(\xi)G(\xi)X
	\qquad{\rm projected\enspace on\enspace}{\cal E}      $$
$$	=D_0(1+aD_0^{-1}D'(\xi)+a^2D_0^{-1}D'(\xi)G(\xi))X
	\qquad{\rm projected\enspace on\enspace}{\cal E}      $$
Let now $(u^{(i)})$ and $(u^{(i)\perp})$ be conjugate bases of ${\cal E}$.  Also let $\delta^{(i)}$ for $i=1,\ldots,m_1+\ldots+m_r$ be the eigenvalues of $D_0$ restricted to ${\cal E}$.  Then, to second order in $a$,
$$	N(\xi,a)\wedge_1^{m_1+\ldots+m_r}u^{(\ell)}      $$
is, up to a factor of absolute value 1,
$$	(\prod_{\ell=1}^{m_1+\ldots+m_r}\delta^{(\ell)})[1+a\sum_{i=1}
^{m_1+\ldots+m_r}(u^{(i)\perp},D_0^{-1}D'(\xi)u^{(i)})      $$
$$	+a^2\sum_{i<j}
((u^{(i)\perp},D_0^{-1}D'(\xi)u^{(i)})(u^{(j)\perp},D_0^{-1}D'(\xi)u^{(j)})  $$
$$	-(u^{(i)\perp},D_0^{-1}D'(\xi)u^{(j)})
	(u^{(j)\perp},D_0^{-1}D'(\xi)u^{(i)})
	+a^2\sum_i(u^{(i)\perp},D_0^{-1}D'(\xi)G(\xi)u^{(i)})]
	\wedge_\ell u^{(\ell)}      $$
so that
$$	N(\xi,a)=(\prod_{\ell=1}^{m_1+\ldots+m_r}|\delta^{(\ell)}|)
	[1+\{a\sum_i(u^{(i)\perp},D_0^{-1}D'(\xi)u^{(i)})      $$
$$	+a^2\sum_{i<j}
((u^{(i)\perp},D_0^{-1}D'(\xi)u^{(i)})(u^{(j)\perp},D_0^{-1}D'(\xi)u^{(j)})  $$
$$ -(u^{(i)\perp},D_0^{-1}D'(\xi)u^{(j)})(u^{(j)\perp},D_0^{-1}D'(\xi)u^{(i)}))
	+a^2\sum_i(u^{(i)\perp},D_0^{-1}D'(\xi)G(\xi)u^{(i)})\}]      $$
Since $\log|\delta^{(\ell)}|=\lambda_0^{(\ell)}$ we obtain, to second order in $a$,
$$	L_a=\int d\xi\,\log N(\xi,a)=m_1\lambda_1+\ldots+m_r\lambda_r+\int d\xi
\,[\{\ldots\}-{a^2\over2}(\sum_i(u^{(i)\perp},D_0^{-1}D'(\xi)u^{(i)}))^2]    $$
where $\{\ldots\}$ has the same meaning as above.  Write 
$$	\Psi_i(\sum_\ell\xi_\ell u^{(\ell)})
	=(u^{(i)\perp},D_0^{-1}F'(\sum_\ell\xi_\ell u^{(\ell)}))      $$
The first term of $\int d\xi\,\{\ldots\}$ is 
$$	a\sum_i\int d\xi\,(u^{(i)\perp},D_0^{-1}TF'(\xi)u^{(i)})
	=a\sum_i\int d\xi\,{\partial\over\partial\xi_i}\Psi_i      $$
which vanishes because $\int d\xi\,{\partial\over\partial\xi_i}\ldots=0$.  The next term in $\int d\xi\,\{\ldots\}$ is
$$	a^2\sum_{i<j}\int d\xi\,
(({\partial\Psi_i\over\partial\xi_i})({\partial\Psi_j\over\partial\xi_j})
-({\partial\Psi_i\over\partial\xi_j})({\partial\Psi_j\over\partial\xi_i}))
	=a^2\sum_{i<j}\int d\xi\,
({\partial\over\partial\xi_i}(\Psi_i{\partial\Psi_j\over\partial\xi_j})
-{\partial\over\partial\xi_j}(\Psi_i{\partial\Psi_j\over\partial\xi_i}))   $$
which vanishes as above.  Thus we are left with
$$	L_a-(m_1\lambda_1+\ldots+m_r\lambda_r)      $$
$$	=a^2\int d\xi\,[\sum_i(u^{(i)\perp},D_0^{-1}D'(\xi)G(\xi)u^{(i)})
-{1\over2}(\sum_i(u^{(i)\perp},D_0^{-1}D'(\xi)u^{(i)}))^2]\eqno{(4)}      $$
and we may write, using (3),
$$	\sum_i(u^{(i)\perp},D_0^{-1}D'(\xi)G(\xi)u^{(i)})
	=-\sum_{n=0}^\infty\sum_i
(u^{(i)\perp},D_0^{-1}D'(\xi)D_0^{-n-1}D'_\perp(F_0^n)D_0^nu^{(i)})      $$
$$	=-\sum_{n=0}^\infty\sum_i\sum^*_j(u^{(i)\perp},D_0^{-1}D'(\xi)u^{(j)})
	(u^{(j)\perp},D_0^{-n-1}D'(F_0^n\xi)D_0^nu^{(i)})      $$
where we have introduces conjugate bases $(u^{(j)})$, $(u^{(j)\perp})$ of ${\cal E}$, indexed by $j=m_1+\ldots+m_r+1,\ldots,m+1$, and $\sum_i$ is over $i\le m_1+\ldots+m_r+1$, $\sum^*_j$ is over $j>m_1+\ldots+m_r+1$.  The above expression is also
$$	=-\sum_{n=0}^\infty\sum_i\sum^*_j{\partial\over\partial\xi_j}
	(u^{(i)\perp},D_0^{-1}F'(\sum_\ell\xi_\ell u^{(\ell)}))
	{\partial\over\partial\xi_i}
	(u^{(j)\perp},D_0^{-n-1}F'(F_0^n\sum_\ell\xi_\ell u^{(\ell)}))      $$
and integration by part gives thus 
$$	\int d\xi\,\sum_i(u^{(i)\perp},D_0^{-1}D'(\xi)G(\xi)u^{(i)})      $$
$$	=-\sum_{n=0}^\infty\int d\xi\,\sum_i{\partial\over\partial\xi_i}
	(u^{(i)\perp},D_0^{-1}F'(\sum_\ell\xi_\ell u^{(\ell)}))
	\sum^*_j{\partial\over\partial\xi_j}
	(u^{(j)\perp},D_0^{-n-1}F'(F_0^n\sum_\ell\xi_\ell u^{(\ell)}))      $$
$$	=-\sum_{n=0}^\infty\int d\xi\,{\rm Tr}_{\cal E}(D_0^{-1}D'(\xi))
	{\rm Tr}_{{\cal E}^\perp}(D_0^{-n-1}D'(F_0^n\xi)D_0^n)      $$
$$	=-\sum_{n=0}^\infty\int d\xi\,{\rm Tr}_{\cal E}(D_0^{-1}D'(\xi))
	{\rm Tr}_{{\cal E}^\perp}(D_0^{-1}D'(F_0^n\xi))      $$
The fact that $F=F_0+aF'$ is volume preserving is expressed by ${\rm Tr}_{{\bf R}^{m+1}}(D_0^{-1}D'(\xi))=0$ hence
$$	\int d\xi\,\sum_i(u^{(i)\perp},D_0^{-1}D'(\xi)G(\xi)u^{(i)})      $$
$$	=\sum_{n=0}^\infty\int d\xi\,{\rm Tr}_{\cal E}(D_0^{-1}D'(\xi))
	{\rm Tr}_{\cal E}(D_0^{-1}D'(F_0^n\xi))      $$
and introducing this in (4) yields
$$	L_a-(m_1\lambda_1+\ldots+m_r\lambda_r)      $$
$$	=a^2[\sum_{n=1}^\infty\int d\xi\,{\rm Tr}_{\cal E}(D_0^{-1}D'(\xi))
	{\rm Tr}_{\cal E}(D_0^{-1}D'(F_0^n\xi))
	+{1\over2}\int d\xi\,({\rm Tr}_{\cal E}(D_0^{-1}D'(\xi)))^2]      $$
$$	={a^2\over2}\sum_{n=-\infty}^\infty\int d\xi\,
	{\rm Tr}_{\cal E}(D_0^{-1}D'(\xi))
	{\rm Tr}_{\cal E}(D_0^{-1}D'(F_0^n\xi))\eqno{(5)}      $$
where the last step used the invariance of $d\xi$ under $F_0^n$.\qed
\medskip
	{\bf 10. Proof of Theorem 1.}
\medskip
	We use Proposition 9, the corresponding result with $F$ replaced by $F^{-1}$, and the fact that $\sum_{\ell=1}^m\lambda_a^{(\ell)}=0$ (because $F$ is volume preserving).  This gives an estimate of all the sums of $\lambda_a^{(\ell)}$ that occur in Theorem 1.\qed
\medskip
	{\bf 11. Corollary.}
\medskip
	{\sl In the situation of Theorem 1, the central Lyapunov exponent is 
$$	\lambda^c={a^2\over2}\sum_{-\infty}^{\infty}\int d\xi\,
	[{\rm Tr}^u(D_0^{-1}D'(\xi)){\rm Tr}^u(D_0^{-1}D'(F_0^n\xi))
	-{\rm Tr}^s(D_0^{-1}D'(\xi)){\rm Tr}^s(D_0^{-1}D'(F_0^n\xi))]      $$
$$	={a^2\over2}\sum_{-\infty}^{\infty}\int d\xi\,
	[{\rm Tr}^s(D_0^{-1}D'(\xi))-{\rm Tr}^u(D_0^{-1}D'(\xi))]
	{\rm Tr}^c(D_0^{-1}D'(F_0^n\xi))      $$
where ${\rm Tr}^s$, ${\rm Tr}^u$, ${\rm Tr}^c$ denote the traces over the spectral subspaces ${\cal E}^s$, ${\cal E}^u$, ${\cal E}^c$ of $D_0$ corresponding to eigenvalues $<1$, $>1$, or $=1$ in absolute value (${\cal E}^c$ is one dimensional).}
\medskip
	Since $F$ preserves the volume, the sum of all Lyapunov exponents vanishes.  Therefore $\lambda^c$ is minus the sum of the negative Lyapunov exponents, given by (5), minus the sum of the positive Lyapunov exponents.  Note that replacing $F$ by $F^{-1}$, ${\cal E}^s$ by ${\cal E}^u$ (and, to the order considered, $D'(\xi)$ by $-D'(\xi)$) replaces the sum of the negative Lyapunov exponents by minus the sum of the positive exponents.  This gives the first formula for $\lambda^c$.
\medskip
	To obtain the second formula, express ${\rm Tr}^u{\rm Tr}^u-{\rm Tr}^s{\rm Tr}^s$ in terms of ${\rm Tr}^u\pm{\rm Tr}^s$, and remember that (because $F$ preserves the volume) ${\rm Tr}^s+{\rm Tr}^u+{\rm Tr}^c=0$ when applied to $D_0^{-1}D'(\xi)$.\qed
\medskip
	The above formula (5) takes a particularly simple form in a special case described in the next theorem.  
\medskip
	{\bf 12. Theorem.}
\medskip
	{\sl Let $\Phi$ be a hyperbolic automorphism of ${\bf T}^m$, with stable and unstable dimensions $m^s$ and $m^u=m-m^s$, and with entropy $\lambda_0^u$.  Let $J:y\to y+\alpha$ {\rm (mod 1)} be a translation of ${\bf T}$, and $\phi:{\bf T}^m\to{\bf T}$ a morphism $\ne0$.  Finally let $\psi:{\bf T}\to{\bf R}^m$ be a nullhomotopic {\rm C}$^2$ function.
\medskip
	Define $h,g_a:{\bf T}^m\times{\bf T}\to{\bf T}^m\times{\bf T}$ by
$$	h\big(\matrix{x\cr y\cr}\big)
	=\big(\matrix{\Phi x\cr Jy+\phi\Phi x-\phi x\cr}\big)\qquad,\qquad
	g_a\big(\matrix{x\cr y\cr}\big)
	=\big(\matrix{x+a\psi(y)\enspace{\rm (mod\enspace1)}\cr y\cr}\big)  $$
and let $f_a=g_a\circ h$.
\medskip
	Denote by $\lambda_a^s$ (resp. $\lambda_a^u$) the sum of the smallest $m^s$ (resp. the largest $m^u$) Lyapunov exponents for $(f_a,{\rm volume})$.  Also let $\lambda_a^c=-\lambda_a^s-\lambda_a^u$ be the ``central exponent''.  Then $\lambda_a^s$, $\lambda_a^u$, $\lambda_a^c$ have expansions of order 2 in $a$:
$$	\lambda_a^s=-\lambda_0^u
	+{a^2\over2}\int_{\bf T}dy\,((\nabla\phi)\psi'^s(y))^2+o(a^2)      $$
$$	\lambda_a^u=\lambda_0^u
	-{a^2\over2}\int_{\bf T}dy\,((\nabla\phi)\psi'^u(y))^2+o(a^2)      $$
$$	\lambda_a^c={a^2\over2}\int_{\bf T}dy\,[((\nabla\phi)\psi'^u(y))^2
	-((\nabla\phi)\psi'^s(y))^2]+o(a^2)      $$
Here $\psi'^s(y)$ and $\psi'^u(y)$ are the components of the derivative $\psi'(y)\in{\bf R}^m$ in the stable and unstable subspaces ${\cal E}^s$ and ${\cal E}^u$ for $\Phi$.  Also, we have used $\nabla\phi:{\bf R}^m\to{\bf R}$ to denote the derivative of the map $\phi:{\bf T}^m\to{\bf T}$ with the obvious identifications.}
\medskip
	This theorem is a simple (but nontrivial) extension of the result proved by Shub and Wilkinson [10].  In the situation that they consider $\Phi=\big(\matrix{2&1\cr1&1\cr}\big)$, $J=$identity, $\phi=(1,0)$, $\psi'=\psi'^u$.  [Remark that, in the notation of [10], $u_0=((1,1).v_0)/(m-1)=((1,0).v_0)$ so that the formula given in Proposition II of [10] agrees with our result above].
\medskip
	{\bf Notation.}
\medskip
	We shall henceforth omit the (mod 1).  We shall keep $\nabla$ to denote the derivative in ${\bf T}^m$.  With obvious abuses of notation, the reader may find it convenient to think of $\Phi$ or $\nabla\Phi$ as an $m\times m$ matrix (with integer entries and determinant $\pm1$), and $\phi$ or $\nabla\phi$ as a row $m$-vector (with integer entries not all zero).  
\medskip
	{\bf 13. Reformulation of the problem.}
\medskip
	Note that $f_a^{-1}=h^{-1}\circ g_a^{-1}$ where $h^{-1}$, $g_a^{-1}$ are obtained from $h$, $g_a$ by the replacements $\Phi,J,\phi,\psi\to\Phi^{-1},J^{-1},\phi,-\psi$.  These replacements also interchange the stable and unstable subspaces for $\Phi$ and replace $\lambda^s$, $\lambda^u$ by $-\lambda^u$, $-\lambda^s$.  Therefore the formula for $\lambda^u$ in the theorem follows from the formula for $\lambda^s$.  And the formula for $\lambda^c=-\lambda^s-\lambda^u$ also follows.  To complete the proof of the theorem we turn now to the formula for $\lambda^s$.
\medskip
	Define
$$ \hat\phi\big(\matrix{x\cr y\cr}\big)=\big(\matrix{x\cr y+\phi x\cr}\big) $$
then
$$	F_0\big(\matrix{x\cr y\cr}\big)
	=\hat\phi^{-1}h\hat\phi\big(\matrix{x\cr y\cr}\big)
	=\big(\matrix{\Phi x\cr Jy\cr}\big)      $$
$$	\hat g_a\big(\matrix{x\cr y\cr}\big)
	=\hat\phi^{-1}g_a\hat\phi\big(\matrix{x\cr y\cr}\big)
=\big(\matrix{x+a\psi(y+\phi x)\cr y-a(\nabla\phi)\psi(y+\phi x)\cr}\big) $$
so that
$$	F\big(\matrix{x\cr y\cr}\big)
	=\hat\phi^{-1}f_a\hat\phi\big(\matrix{x\cr y\cr}\big)
	=\hat g_a F_0\big(\matrix{x\cr y\cr}\big)
	=\big(\matrix{\Phi x+a\psi(Jy+\phi\Phi x)\cr 
	Jy-a(\nabla\phi)\psi(Jy+\phi\Phi x)\cr}\big)      $$
Finally, $F=F_0+aF'$ with
$$	F_0\big(\matrix{x\cr y\cr}\big)
	=\big(\matrix{\Phi x\cr Jy\cr}\big)\qquad,\qquad
	F'\big(\matrix{x\cr y\cr}\big)=\big(\matrix{\psi(Jy+\phi\Phi x)\cr 
	-(\nabla\phi)\psi(Jy+\phi\Phi x)\cr}\big)      $$
Since $F$ is conjugate (linearly) to $f_a$, we may compute $\lambda^s$ from $F$ instead of $f_a$.\medskip
	{\bf 14. Proof of Theorem 12.}
\medskip
	Write ${\bf R}^{m+1}={\cal E}^s+{\cal E}^u+{\bf R}$.  We shall apply Proposition 9 with ${\cal E}={\cal E}^s$, ${\cal E}^\perp={\cal E}^u+{\bf R}$.  Using $\xi=(x,y)$ and $X\in{\cal E}^s$, $Y\in{\cal E}^u$, $Z\in{\bf R}$ we may write
$$	D_0\big(\matrix{X+Y\cr Z\cr}\big)
	=\big(\matrix{(\nabla\Phi)(X+Y)\cr Z\cr}\big)      $$
$$	D'(\xi)\big(\matrix{X+Y\cr Z\cr}\big)
	=\big(\matrix{\psi'(Jy+\phi\Phi x)((\nabla\phi\Phi)(X+Y)+Z)\cr
-(\nabla\phi)\psi'(Jy+\phi\Phi x)((\nabla\phi\Phi)(X+Y)+Z)\cr}\big)      $$
where $\psi'$ denotes the derivative of $\psi$.  Therefore 
$$  {\rm Tr}_{\cal E}(D'(\xi)D_0^{-1})=(\nabla\phi)\psi'^s(Jy+\phi\Phi x)  $$
and (5)contains the integrals
$$	\int d\xi\,{\rm Tr}_{\cal E}(D_0^{-1}D'(\xi))
	{\rm Tr}_{\cal E}(D_0^{-1}D'(F_0^n\xi))      $$
$$	=\int d\xi\,[(\nabla\phi)\psi'^s(Jy+\phi\Phi x)]
	[(\nabla\phi)\psi'^s(J^{n+1}y+\phi\Phi^{n+1}x)]      $$
Performing a change of variables $\bar x=\Phi x$, $\bar y=Jy+\phi\Phi x$ we find that this is 
$$	=\int d\bar x\,d\bar y\,[(\nabla\phi)\psi'^s(\bar y)]
	[(\nabla\phi)\psi'^s(J^n\bar y+\phi\Phi^n\bar x-\phi\bar x)]      $$
We claim that this last integral vanishes unless $n=0$.  This is because, if $n\ne0$, 
$$  \int d\bar x\,\psi'(J^n\bar y+\phi\Phi^n\bar x-\phi\bar x)=0  $$
Indeed, $\phi\Phi^n\bar x-\phi\bar x$ is a linear combination with integer coefficients of the components $\bar x_1,\ldots,\bar x_m$ of $\bar x$, and the coefficients do not all vanish because $\phi\Phi^n=\phi$ is impossible ($\Phi$ is hyperbolic and $\phi\ne0$).  Integrating the derivative $\psi'$ with respect to a variable $\bar x_j$ really occuring in $\phi\Phi^\ell\bar x-\phi\bar x$ gives zero as announced.  
\medskip
	Returning to (5) we have thus 
$$	\lambda_a^s+\lambda_o^u
	={a^2\over2}\int d\xi\,({\rm Tr}_{\cal E}(D_0^{-1}D'(\xi)))^2      $$
$$	={a^2\over2}\int d\bar y((\nabla\phi)\psi'^s(\bar y))^2      $$
which is the formula given for $\lambda_0^s$ in Theorem 12.  And according to Section 13 this completes our proof.\qed
\medskip
	{\bf 15. Final remarks.}
\medskip
	(a) Shub and Wilkinson [10] showed that close to a diffeomorphism $($hyperbolic automorphism $\Phi$ of ${\bf T}^2)\times($identity on ${\bf T})$ there is a C$^1$ open set of ergodic volume preserving C$^2$ diffeomorphisms of ${\bf T}^3$ with central Lyapunov exponent $\lambda^c>0$.  They remark that their result extends to the situation where $\Phi$ is a hyperbolic automorphism of ${\bf T}^m$ with one-dimensional expanding eigenspace.  More generally, if $\Phi$ is any hyperbolic automorphism of ${\bf T}^m$, Theorem 12 gives close to $(\Phi,$ rotation of ${\bf T})$ in C$^2({\bf T}^{m+1})$ a diffeomorphism $F$ with $\lambda^c>0$.  Since $\lambda^c$ is given by an integral over the volume of a local ``central'' stretching exponent, we have $\lambda^c>0$ in a C$^1$ neighborhood of $F$.  But by a result of Dolgopyat and Wilkinson [5] (Corollary 0.5), stable ergodicity is here C$^1$ open and dense in the C$^2$ volume preserving diffeomorphisms: we have center bunching and stable dynamical coherence because we consider perturbations of $(\Phi,$ rotation of ${\bf T})$ for which the center foliation is C$^1$, see [6], [7].  In conclusion, close to $($hyperbolic automorphism $\Phi$ of ${\bf T}^m)\times($rotation on ${\bf T})$ there is a C$^1$ open set $V$ of ergodic volume preserving C$^2$ diffeomorphisms of ${\bf T}^{m+1}$ with central Lyapunov exponent $\lambda^c>0$ (or also with $\lambda^c<0$).  In particular, if $F\in V$, the conditional measures of the volume on the circles $\pi^{-1}\{x\}$ are atomic, as discussed in [9].
\medskip
	(b) The coefficient $L$ in Proposition 9 is $\ge0$.  Consider indeed the unitary operator $U$ defined by $U\psi=\psi\circ F$ on $L^2({\bf T}^{m+1},{\rm volume})$, and let $E(.)$ be the corresponding spectral measure, so that 
$$	U=\int_{\bf T}e^{2\pi i\theta}E(d\theta)      $$
If $\psi(\xi)={\rm Tr}_{\cal E}(D_0^{-1}D'(\xi))$ we have a measure $\nu\ge0$ on ${\bf T}$ defined by $\nu(d\theta)=(\psi,E(d\theta)\psi)$ and the Fourier coefficients
$$	c_n=\int e^{2\pi ni\theta}\nu(d\theta)
=\int d\xi\,{\rm Tr}_{\cal E}(D_0^{-1}D'(\xi))(D_0^{-1}D'(F_0^n\xi))      $$
of this measure tend to zero exponentially.  Therefore $\nu(d\theta)=\rho(\theta)d\theta$ has a smooth density $\rho$ and 
$$	L={1\over2}\sum_{n=-\infty}^\infty c_n={1\over2}\rho(0)\ge0      $$
\medskip
	(c) Suppose now that $F$ is not necessarily a volume preserving perturbation of $F_0$.  We may still hope that $F$ has an SRB measure $\rho_a$.  If $F_0$ were hyperbolic, we would have an expansion
$$	\rho_a=\rho_0+a\delta+o(a)      $$
(see [8]) with $\rho_0=$ Lebesgue measure and $\delta$ a distribution.  For smooth $\Psi$, $\delta(\Psi)$ is given (because $\rho_0$ is Lebesgue measure) by the simple formula (see [8])
$$	\delta(\Psi)
=-\sum_0^\infty\rho_0((\Psi\circ F_0^n).{\rm div}(F'\circ F_0^{-1})      $$
Similarly (replacing $F$ by $F^{-1}$, hence $F_0$, $D_0^{-1}D'(\xi)$ by $F_0^{-1}$, $-D'(F_0^{-1}\xi)D_0^{-1}$ we see that the anti-SRB state has an expansion 
$$	\bar\rho_a=\rho_0+a\bar\delta+o(a)      $$
with 
$$	\bar\delta(\Psi)=\sum_{n=1}^\infty\int d\xi\,\Psi(F_0^{-n}\xi)
	{\rm Tr}_{{\bf R}^{m+1}}(D'(F_0^{-1}\xi)D_0^{-1})      $$
$$	=\sum_{n=0}^\infty\int d\xi\,\Psi(F_0^{-n}\xi)
	{\rm Tr}_{{\bf R}^{m+1}}(D_0^{-1}D'(\xi))      $$
\medskip
	We can now estimate the Lyapunov exponents for $(F,\rho_a)$ to second order in $a$ even though we are not sure of the existence of the SRB measure $\rho_a$.  We simply assume that we can use the formula for $\delta(\Psi)$.  Going through the proof of Proposition 9 we have to replace $\int d\xi\,\log N(\xi,a)$ by $\rho_a(\log N(.,a))$ and (to second order in $a$) this adds to the right-hand side of (4) a term
$$	-a^2\sum_{n=1}^\infty\int d\xi\,{\rm Tr}_{\cal E}(D_0^{-1}D'(\xi))
	{\rm Tr}_{{\bf R}^{m+1}}(D_0^{-1}D'(\xi))      $$
Taking into account the integrations by part we obtain now instead of (5) the formula
$$	L_a-(m_1\lambda_1+\dots+m_r\lambda_r)
	={a^2\over2}\sum_{n=-\infty}^\infty\int d\xi\,
	{\rm Tr}_{\cal E}(D_0^{-1}D'(\xi))
	{\rm Tr}_{\cal E}(D_0^{-1}D'(F_0^n\xi))      $$
$$	-a^2\sum_{n=-\infty}^\infty\int d\xi\,
	{\rm Tr}_{\cal E}(D_0^{-1}D'(\xi))
	{\rm Tr}_{{\bf R}^{m+1}}(D_0^{-1}D'(F_0^n\xi))\eqno{(6)}      $$
\indent
	Let $a^2L^s$, $a^2L^u$, $a^2L^c$ be the $a^2$ contributions to the sum of the noncentral negative, noncentral positive, and the central Lyapunov exponents for the SRB measure.  We obtain $a^2L^s$ from (6) when $n_r=n^s$.  A similar calculation gives $a^2L^u$ (it is convenient here to work via the anti-SRB measure, then replace $F$ by $F^{-1}$).  Estimating the average expansion coefficient gives $a^2(L^s+L^u+L^c)=\rho_a(\log\det(D_0+aD'(.))$, hence $L^s+L^u+L^c$, hence $L^c$.  The results are
$$	L^s={1\over2}\sum_{n=-\infty}^\infty\int d\xi\,
	{\rm Tr}^s(D_0^{-1}D'(\xi)){\rm Tr}^s(D_0^{-1}D'(F_0^n\xi))      $$
$$	-\sum_{n=-\infty}^\infty\int d\xi\,
	{\rm Tr}^s(D_0^{-1}D'(\xi))
	{\rm Tr}_{{\bf R}^{m+1}}(D_0^{-1}D'(F_0^n\xi))      $$
$$	L^u=-{1\over2}\sum_{n=-\infty}^\infty\int d\xi\,
	{\rm Tr}^u(D_0^{-1}D'(\xi)){\rm Tr}^u(D_0^{-1}D'(F_0^n\xi))      $$
$$	L^c=-{1\over2}\sum_{n=-\infty}^\infty\int d\xi\,
	{\rm Tr}^c(D_0^{-1}D'(\xi)){\rm Tr}^c(D_0^{-1}D'(F_0^n\xi))      $$
$$	-\sum_{n=-\infty}^\infty\int d\xi\,
	{\rm Tr}^c(D_0^{-1}D'(\xi)){\rm Tr}^u(D_0^{-1}D'(F_0^n\xi))      $$
$$	L^s+L^u+L^c=-{1\over2}\sum_{n=-\infty}^\infty\int d\xi\,
	{\rm Tr}_{{\bf R}^{m+1}}(D_0^{-1}D'(\xi))
	{\rm Tr}_{{\bf R}^{m+1}}(D_0^{-1}D'(F_0^n\xi))      $$
which can be rewritten variously.
\medskip
	The existence of second order expansions for the Lyapunov exponents gives added interest to the question whether the SRB measure $\rho_a$ really exists for (small) finite $a$.
\medskip
	(d) The author has not looked seriously into possible extensions of the results presented here.  Generalizations are thus left for the reader to formulate, and to prove.
\vfill\eject
	{\bf References.}  
\medskip
	[1] A. Avila and J. Bochi.  ``A formula with some applications to the theory of Lyapunov exponents.''  Preprint

	[2] C. Bonatti, X. G\'omez-Mont and M. Viana.  ``G\'en\'ericit\'e d'expoants de Lyapunov non-nuls des produits d\'eterministes de matrices.''  Preprint

	[3] K. Burns and A. Wilkinson.  ``Stable ergodicity of skew products.''  Ann. Sci. Ecole Norm. Sup. {\bf 32},859-889(1999).

	[4] D. Dolgopyat.  ``On differentiability of SRB states.''  Preprint.

	[5] D. Dolgopyat and  A. Wilkinson.  ``Stable accessibility is C$^1$ dense.''  Preprint

	[6] M. Hirsch, C.C. Pugh, and M. Shub.  {\it Invariant manifolds.}  Lect. Notes in Math. {\bf 583} Springer, Berlin, 1977.

	[7] C. Pugh, and M. Shub.  ``Stable ergodicity and julienne quasi-conformality.''  J. Eur. Math. Soc. {\bf 2},1-52(2000).

	[8] D. Ruelle.  ``Differentiation of SRB states.''  Commun. Math. Phys. {\bf 187},227-241(1997). 

	[9] D. Ruelle and  A. Wilkinson.  ``Absolutely singular dynamical foliations.''  Commun. Math. Phys., to appear.

	[10] M. Shub and  A. Wilkinson.  ``Pathological foliations and removable exponents.''  Inventiones Math. {\bf 139},495-508(2000).
\end